\documentclass[%
 aip,
 amsmath,amssymb,
 reprint,%
]{revtex4-1}

\usepackage{graphicx}% Include figure files
\usepackage{dcolumn}% Align table columns on decimal point
\usepackage{bm}% bold math
\usepackage[utf8]{inputenc}
\usepackage[T1]{fontenc}
\usepackage{mathptmx}
\usepackage{etoolbox}

\usepackage{subcaption} %Para generar subfigure
\usepackage[normalem]{ulem} % aux: para tachar \sout
\usepackage{comment} %Para comentar fragmentos
\usepackage{color,soul} %highlight \hl{TEXTO}
\usepackage{hyperref} %Para hipervínculos 

\usepackage{ragged2e} % alineaciones de texto
\makeatletter
\def\@email#1#2{%
 \endgroup
 \patchcmd{\titleblock@produce}
  {\frontmatter@RRAPformat}
  {\frontmatter@RRAPformat{\produce@RRAP{*#1\href{mailto:#2}{#2}}}\frontmatter@RRAPformat}
  {}{}
}%
\makeatother
\begin{document}

\preprint{AIP/123-QED}

\title[Effectiveness of Turbulent Fountains]{Effectiveness of Turbulent Fountains in Frost Mitigation and Pollution Control}

% Force line breaks with \\

\author{Daniel Freire Caporale}
\email{dfreire@fisica.edu.uy}
\affiliation{Instituto de Física, Universidad de la República, Uruguay.}%
%\homepage{http://fisicanolineal.fisica.edu.uy/}

\author{Luis G. Sarasúa}
\affiliation{Instituto de Física, Universidad de la República, Uruguay.}

\author{Nicasio Barrere}
\affiliation{Centro Universitario del Este, Universidad de la República, Uruguay.}%

\author{Arturo C, Martí}%
\affiliation{Instituto de Física, Universidad de la República, Uruguay.}
%\affiliation{Authors' institution and/or address%\\This line break forced with \textbackslash\textbackslash
%}%

\date{\today}% It is always \today, today,

\begin{abstract}
%This study investigates the efficacy of turbulent fountains as a method for frost mitigation and pollution control in stratified ambient fluids. By analysing the interaction between the ejected fountain fluid and the lower layers of the atmosphere, we develop predictive tools to assess the impact on temperature distribution and the amount of ejected fluid, considered polluted, that returns to the ground. This is evaluated for various fountain configurations, $\left(Fr^{-2}, u^{\prime}/U\right)$, where $Fr$ is the Froude number and $u^{\prime}/U$ is the turbulent intensity of the fountain at the inlet.
%
%Notably, an optimal range in the configuration space exists, where the temperature increase is maximised, regardless of the fountain’s turbulence intensity. This optimal region corresponds to the so-called semi-collapse regime of the fountain, as described by Sarasúa et al. (2024). In this zone, the returning fluid, having mixed with warmer upper ambient layers, contributes to a significant temperature rise. Our analysis highlights the delicate balance between fountain conditions and their effects on the surrounding ambient fluid, providing insights into the practical application of turbulent fountains for environmental control.
We investigate the efficacy of turbulent fountains as a tool for frost mitigation and pollution control in stratified ambient fluids. By examining the interaction between the ejected fountain fluid and the lower atmospheric layers, we develop predictive models to assess their impact on temperature distribution and the return of polluted ejected fluid to the ground. This analysis covers a range of fountain parameters, including $Fr^{-2}$ and $u^{\prime}/U$, where $Fr$ is the Froude number and $u^{\prime}/U$ represents the turbulent intensity at the inlet. Notably, we identify an optimal set of parameter values where the temperature rise is maximised, independent of the fountain’s turbulence intensity. This optimal condition occurs in the so-called semi-collapse regime, as described by Sarasúa et al. [Flow, Turbulence and Combustion, 112(4), 1009-1025 (2024)], where the returning fluid, mixed with warmer upper ambient layers, significantly increases local temperature. Our findings underscore the importance of carefully tuning fountain parameters to balance their effects on the surrounding environment, offering valuable insights for the practical use of turbulent fountains in environmental management.

\end{abstract}

\maketitle

%\begin{quotation}
%The ``lead paragraph'' is encapsulated with the \LaTeX\ 
%\verb+quotation+ environment and is formatted as a single paragraph before the first section heading. 
%(The \verb+quotation+ environment reverts to its usual meaning after the first sectioning command.) 
%Note that numbered references are allowed in the lead paragraph.
%
%The lead paragraph will only be found in an article being prepared for the journal \textit{Chaos}.
%\end{quotation}

%%%%%%%%%%%%%%%%%%%%%%%%%%%%

\section{\label{sec:intro}Introduction}

A fountain, defined as a vertical buoyant jet where the buoyancy force opposes the jet's initial velocity, contrasts with a plume, in which buoyancy and velocity are aligned. Both fountains and plumes are ubiquitous in natural systems and technological applications, particularly where the dynamics of stratified fluids are of interest. Research into turbulent fountains and plumes in both homogeneous and stratified environments has spanned several decades (e.g., Morton et al. 1956\cite{morton1956turbulent}; Kaye 2008\cite{Kaye}; Woods 2010\cite{Woods}), with early studies focusing on turbulent jets due to their relevance in atmospheric science and environmental engineering.

\begin{figure}[htb]
	\centering
	\includegraphics[width=0.7\columnwidth]{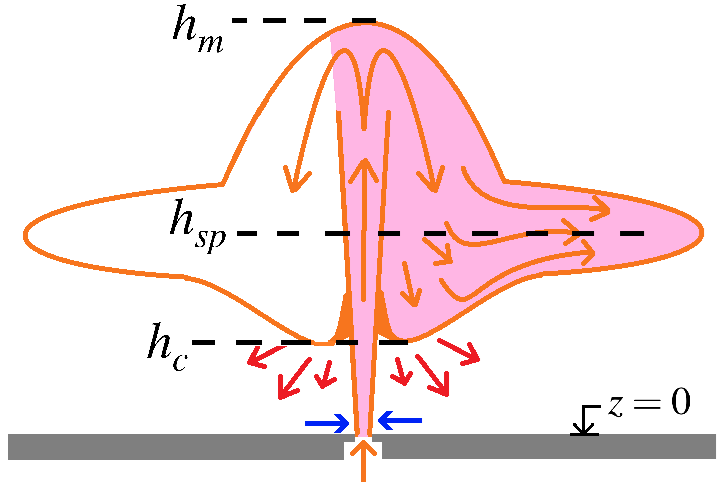}

	\caption{\justifying Two-dimensional schematic of the fountain flow and the characteristic heights of its dynamics in the developed flow regime: the maximum height ($h_m$) attained by the fountain, the spreading height ($h_{sp}$), around which the fountain spreads radially away from the axis in the form of a ``spreading cloud'', and the minimal height ($h_c$), defining the lower boundary of the spreading cloud. The region occupied by the fountain % , where $\Phi > 0$,
    is highlighted in purple, representing fluid either injected or entrained from the surrounding ambient. Arrows indicate the motion of fluid: orange arrows represent fluid from the fountain, blue arrows correspond to the entrainment of ambient fluid near the fountain’s inlet and red arrows, below the $h_c$ level, depict the movement of ambient fluid displaced downward during the formation of the spreading cloud.}
	\label{fig:esuemaFuente}
\end{figure}

Fountain dynamics in stratified media exhibit several key features. As the fountain rises, it decelerates due to the combined effects of ambient fluid entrainment and opposing buoyancy, eventually reaching a maximum height, $h_m$, where the vertical momentum vanishes. The fluid then reverses direction, descending as an annular plume and spreading outward to form a ``spreading cloud'' at a height $h_{sp}$ above the fountain ($h_{sp}>0$), or collapsing back to the ground ($h_{sp}\simeq 0$)~\cite{sarasua2024influence}. This spreading or collapse depends strongly on the initial momentum, buoyancy flux, and the stratification profile of the surrounding medium. In Fig.~\ref{fig:esuemaFuente}, we show a schematic of the flow in a no-collapse scenario, highlighting the characteristic flow heights. Additionally, the minimal height, $h_c$, defined as the minimum height of the lower boundary of the spreading cloud and introduced in Sarasúa et al. (2024)~\cite{sarasua2024influence}, is also indicated. In orange, we outline the path of the fountain fluid, from injection to radial movement within the spreading cloud, where the fluid moves away from the fountain axis at height $h_{sp}$. In blue, the entrainment of ambient fluid by the fountain is shown, and in red, the movement of displaced ambient fluid during the formation and stabilisation of the spreading cloud is indicated.

\begin{figure}[htb]
	\centering
	\includegraphics[width=0.8\columnwidth]{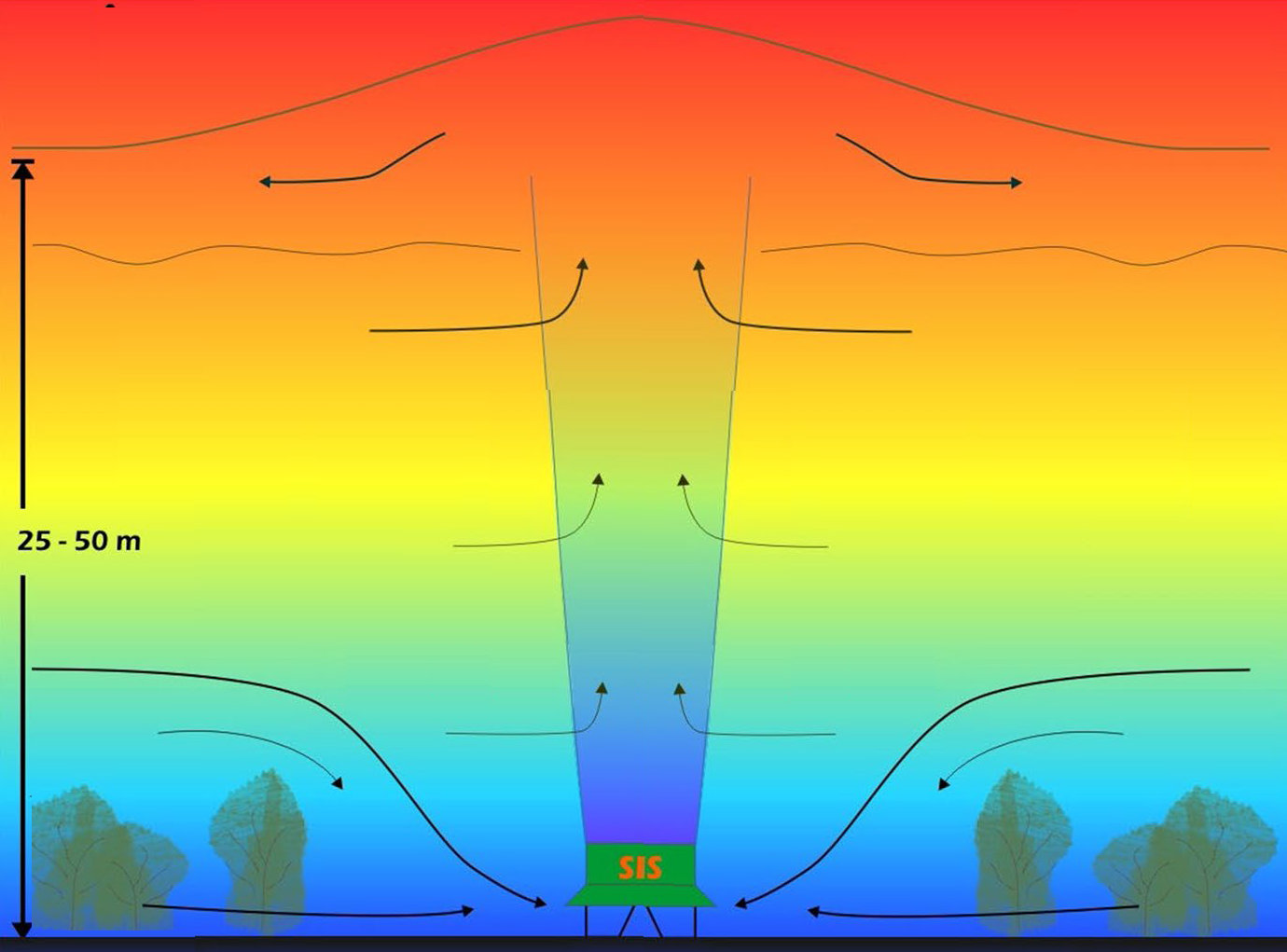}
	\caption{\justifying SIS operation under radiation frost conditions. The linear temperature stratification of the atmosphere is represented by a colour gradient from blue (cold) to red (warm). In a typical operation scenario, $h_{sp}$ ranges from 25 to 50~m. (Illustration adapted with permission from Frost Protection Co. from its website~\cite{SIS_image}.)}
	\label{fig:esquemaSIS}
\end{figure}

As a practical application of fountain dynamics, we refer an innovative technology aimed at protecting crops from frost damage, operating under conditions where no collapse occurs. Frost control in agriculture is critically important, as crop damage depends on both temperature and the duration of exposure, with severe damage occurring at temperatures below $-2 ~^{\circ}\mathrm{C}$~\cite{burke1976freezing,pearce2001plant}.
During radiation frost events, the Earth's surface cools predominantly due to radiative heat loss, leading to the formation of a temperature inversion where the air near the ground becomes the densest and coolest within the ambient environment. This creates an atmospheric stratification characterised by a linear temperature profile with height, as illustrated in Fig.~\ref{fig:esquemaSIS}.
An effective technological solution to this problem is the Selective Inverted Sink (SIS)~\cite{FPC_web} (Guarga et al. 2000~\cite{guarga2000evaluation}).
%\url{https://frostprotection.com/index.php?lang=en-us}).
The SIS works by ejecting cold air upwards, displacing the densest layer of cold air near the ground. This process also entrains air from the surrounding atmosphere (Albertson et al. 1950\cite{albertson1950diffusion}).

A better understanding of these fountain dynamics could further enhance frost protection strategies in agriculture (Arias et al. 2007\cite{arias2007two}; Yazdanpanah and Stigter 2011\cite{yazdanpanah2011selective}; Vahid et al. 2015\cite{vahid2015micrometeorological}; Hu et al. 2018\cite{hu2018review}). Figure~\ref{fig:esquemaSIS} illustrates the core mechanism of the SIS: a vertical-axis propeller draws cold, dense air from the surface through an inlet and expels it upwards through the outlet. Since the SIS operates under conditions that prevent fountain collapse, the cold air does not return to the surface.
%(see \href{https://frostprotection.com/index.php?option=com_content&view=article&id=51&Itemid=466&lang=en-us}).
This upward removal of cold air allows warmer air layers above to descend, thus preventing harmful temperatures near the surface that could damage crops. Following this approach, further applications of the SIS have been proposed for controlling contaminants such as odours or dust in open-pit mining environments~\cite{FPC_services}.
%(see \href{https://frostprotection.com/index.php?option=com_content&view=article&id=49&Itemid=465&lang=en-us}).

The seminal work by Morton et al. (1956)\cite{morton1956turbulent} introduced the foundational \textit{MTT} model, which describes the evolution of volume, momentum, and buoyancy fluxes in fountains. While this model effectively predicts the maximum height in uniform ambient fluids, it does not fully capture post-reversal dynamics or spreading behaviour in stratified environments. Bloomfield and Kerr (1998)\cite{Bloomfield1998} sought to address these limitations by relating the spreading height to the point where fluid densities match. However, their model overlooked the important role of mixing between the fountain and ambient fluid during the downflow phase.

Subsequent studies, such as those by Kaminski et al. (2005)\cite{Kaminski}, refined the understanding of turbulent entrainment by introducing experimentally determined entrainment coefficients, while Mehaddi et al. (2011)\cite{Mehaddi} examined maximum heights in stratified environments. Papanicolaou and Stamoulis (2010)\cite{Papanicolaou} further explored collapse and spreading phenomena. Several studies (e.g., van Reeuwijk and Craske 2015\cite{Reeuwijk}) emphasised the variability of the entrainment coefficient with turbulence intensity, a crucial factor in fountain behaviour.
Recent advances include the work of Sarasúa et al. (2021)\cite{sarasua2021spreading}, who extended the \textit{MTT} framework by incorporating additional parameters, such as the turbulent entrainment coefficient and buoyancy frequency, to more accurately predict collapse and spreading behaviour in axisymmetric turbulent fountains. Freire Caporale et al. (2022)\cite{2022FreireFTLE} further advanced the field by examining Lagrangian coherent structures in the downflow region, using finite-time Lyapunov exponents to quantify entrainment and re-entrainment, thereby improving the understanding of the mechanisms governing fountain dynamics. Freire et al. (2010)\cite{Freire2010} previously investigated the influence of fluctuations on fountain heights $h_m$ and $h_{sp}$, while Sarasúa et al. (2024)\cite{sarasua2024influence} developed a diagram summarising these dependencies, providing practical insights for industrial applications of turbulent fountains. The latter study focused on the collapse of turbulent fountains, analysing the impact of turbulence intensity and the Froude number, two key parameters regulating fountain dynamics. These authors identified three distinct regimes based on $h_{sp}$ and $h_c$, indicated in Fig.~\ref{fig:esuemaFuente}: the collapse regime, where $h_{sp}=0$; the semi-collapse regime, where $h_{sp}>0$ and $h_c=0$; and the no-collapse regime, where both $h_{sp}>0$ and $h_c>0$. Based on this analysis, the authors developed a diagram that divides the configuration space of the fountain (turbulence intensity and Froude number) into regions where each flow regime occurs. 

In this work, we focus on analysing the effect of the fountain flow on the surrounding ambient fluid. While Sarasúa et al. (2024)\cite{sarasua2024influence} investigate the characteristics and flow regime of the fountain based on inlet conditions, the impact on the ambient fluid is not addressed. In this study, we concentrate on that aspect, examining the evolution of the temperature in the lower layers of the ambient fluid (where the SIS operates to increase it and thus protect crops), as well as the concentration of ejected fountain fluid that returns to the ground and the radial distance from the fountain axis where this returning fluid is detected. If we consider that the ejected fluid contains contaminants, it becomes crucial to determine whether it has been entirely removed or what proportion still remains detectable at ground level. Understanding the impact of the fountain flow on the surrounding environment provides a significant contribution to the technological potential of turbulent fountains, aiding in the identification of optimal application conditions.

The paper is organised as follows. In Sec.~\ref{subsec:metodologia}, we detail the methodology employed, which consists of computational simulations validated through laboratory experiments previously\cite{sarasua2021spreading, 2022FreireFTLE} conducted. Then, in Sec.~\ref{sec:indicators}, we present the results along with their analysis, focusing on the impact of the fountain's flow on the temperature of the lower fluid layers and the return of ejected fluid, as a function of the fountain’s characteristics. Finally, in Sec.~\ref{sec:conc}, we provide the conclusion and outline some potential directions for future work.

%Motivación, estado del arte, aportes del trabajo (remoción de contaminantes). Índice.

%Este paper busca aporte sobre el potencial tecnológico de las fuentes turbulentas: Hablar del paper de semi-collapse y decir que hasta allí se vio cuándo el contaminante removido volvía a tomar contacto con el suelo, pero no con qué impacto (en qué cantidad). Tampoco se evaluó el efecto y alcance (radial y en altura) de la fuente sobre el ambiente (aumento/disminución temperatura).

%%%%%%%%%%%%%%%%%%%%%%%%%%%%

\section{\label{subsec:metodologia}Methodology}
%\textcolor{red}{En esta sección, se podría agregar una tablita con configuraciones usadas?  porque dice el rango de u'/U pero luego en resultados usa valores específicos y no queda claro si hay más configs intermedias.  Todas las numéricas tienen validación experimental? no digo de agregar la validación de cada una pero capaz que diría que todos las configuraciones numéricas fueron validadas experimentalmente.}

In this work, the analysis was conducted based on computer simulations using the open-source Computational Fluid Dynamics (CFD) package caffa3d.MBRi~\cite{usera2008parallel}, which implements a fully implicit second-order scheme in both time and space, supporting curvilinear meshes and following the Finite Volume Method (FVM)~\cite{ferziger2020finite}. The numerical results were validated against previously conducted laboratory experiments~\cite{sarasua2021spreading,2022FreireFTLE}, whose setup was designed to rescale the operational conditions of the SIS. This rescaling was achieved through an analysis of the relevant non-dimensional numbers.

As in the laboratory-scale implementation of the SIS’s real-world application, a prismatic computational domain measuring $40~\mathrm{cm} \times 40~\mathrm{cm} \times 50~\mathrm{cm}$ in width ($x$), depth ($y$), and height ($z$) was employed. The working fluid for both the ambient and the fountain was water. The simulations used a time step of $0.05$~s, and the domain was discretised with a computational mesh of $8\times10^{6}$ cells. Mesh independence was confirmed, as no significant differences were observed when using either a coarser mesh of $2\times10^{6}$ cells or a finer mesh of $30\times10^{6}$ cells.

The initial stratification of the ambient fluid follows a linear temperature profile, given by $T_0(z)=T(x,~y,~z,~t = 0)=\partial_z T \cdot z + T_{cold}$, where $T_{cold}=15~^{\circ}\mathrm{C}$ and $\partial_z T = 25~^{\circ}\mathrm{C}\cdot \mathrm{m}^{-1}$, i.e., $T_{hot}=T(x,~y,~z = 50~\mathrm{cm},~t) = 27.5~^{\circ}\mathrm{C}$. Under these conditions, the corresponding characteristic Brünt-Väisälä frequency is $N=0.232~\mathrm{s}^{-1}$. The fountain is injected vertically at temperature $T_{jet}$ through a small circular nozzle with a diameter of $D=0.8$~cm, located at the centre of the base of the domain at $x=y=0$. Calculations were repeated for $T_{jet}=4.0$, 6.0, 7.5, 10.0, 12.5 and 15.0~$^{\circ}\mathrm{C}$.

The boundary conditions of the domain are wall boundaries on all sides, except for the inlet condition at the centre of the base and the outlet condition through a small nozzle at the centre of the upper boundary, with a diameter of $4.0$~cm, where a fully developed flow exit condition was applied. The boundary conditions for temperature are adiabatic on the lateral walls, with constant temperature at the lower boundary $T(x,~y,~z=0,~t)=T_{cold}$, and $T(x,~y,~z=50~\mathrm{cm},~t)=T_{hot}$ on the upper boundary.

At each time step, the fluid velocity in the computational cells at the inlet is given by $\mathbf{U^{in}}=(u_x,~u_y,~u_z+U)$, where $u_x$, $u_y$, and $u_z$ are randomly assigned within the interval $\left[-u^{\prime}, +u^{\prime}\right]$ at each time step, following a uniform distribution, and $U=\dot{q_{in}}/\pi D^2$, where $\dot{q_{in}}=5.5\mathrm{cm}^3\cdot\mathrm{s}^{-1}$ is the fountain's flowrate at the inlet. In this work, for each value of $T_{jet}$, calculations were repeated for different values of $u^{\prime}$ to vary the turbulence intensity $u^{\prime}/U$, taking the values 0, 1, 2, 4, 10 and $20~\%$. Note that at each time step, $u_z+U$ is corrected by a multiplicative factor to ensure that the average over the computational cells at the fountain inlet equals $U$. The turbulence model used is the standard Smagorinsky large-eddy model, with a Smagorinsky coefficient for subgrid viscosity of $C_S=0.16$ in our simulations. Due to the random nature of the imposed fluctuations, simulations were repeated under the same fountain configuration for several cases, yielding similar results.

\begin{figure}[htb]
	\centering
	\includegraphics[width=0.8\columnwidth]{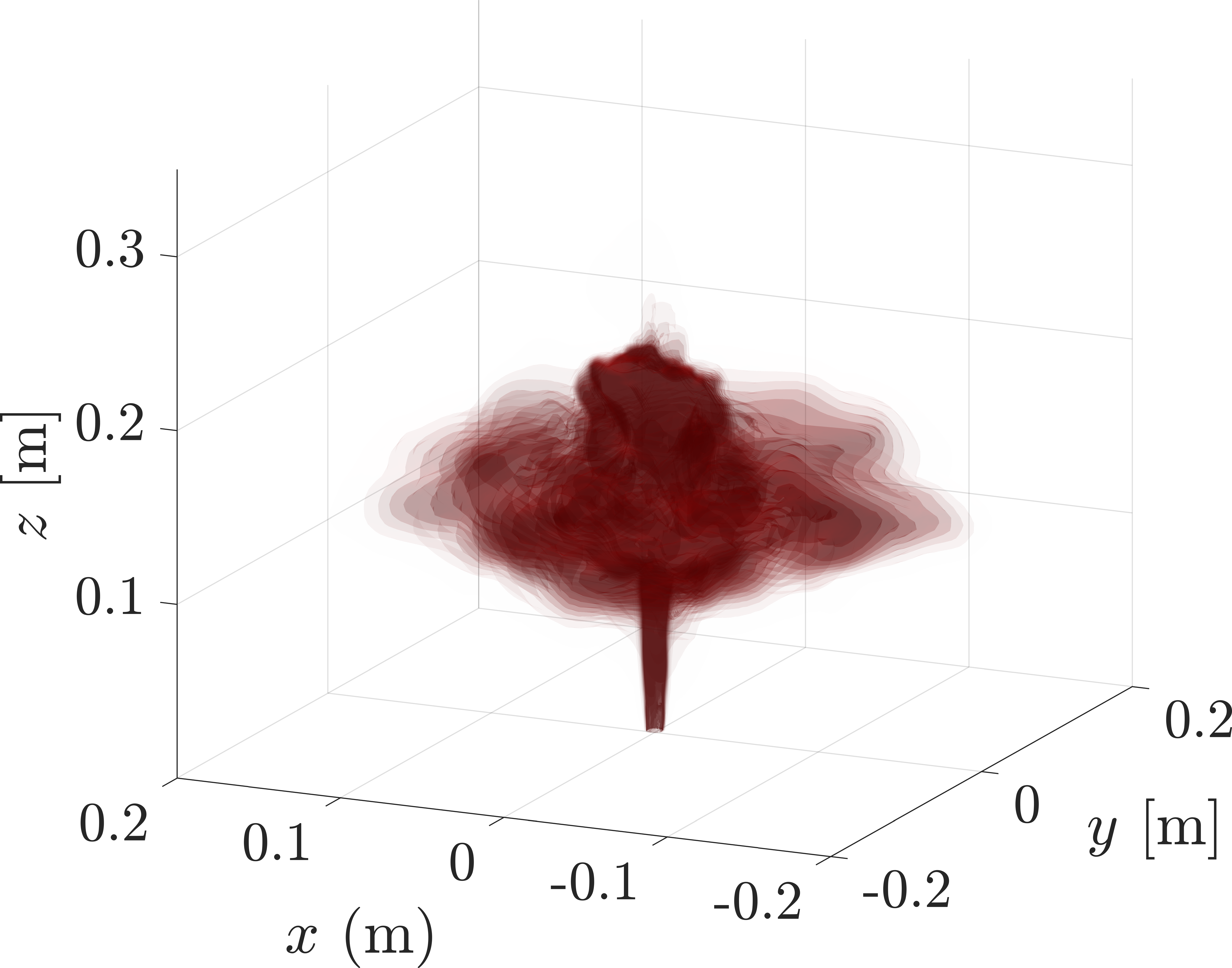}
	\caption{\justifying Three-dimensional view of the passive scalar field $\Phi$, represented in red shades, corresponding to values of $\Phi$ between 0 (white) and 1 (red). This case corresponds to a fountain temperature of $T_{jet}=15~^{\circ}\mathrm{C}$ and a turbulence intensity of $u^{\prime}/U=20~\%$ at $t=80~\mathrm{s}$, when the flow is fully developed.}
	\label{fig:TrazadorSims3D}
\end{figure}

In this work, we analyse the temperature field to study its evolution in the lower layers of the ambient fluid. Additionally, to investigate the mixing phenomenon, a passive scalar field, represented by $\Phi(x,y,z,t)$, was used to track the concentration of injected fluid. Initially, $\Phi(x,y,z,t=0)=0$ throughout the domain, except in the cells located at the inlet nozzle, where $\Phi=1$ at all times. Thus, advection, entrainment, mixing, and diffusion cause $\Phi(x,y,z,t)$ to take values between 0 and 1 throughout the domain, representing the concentration of fluid injected by the fountain. Figure.~\ref{fig:TrazadorSims3D} shows a typical computationally obtained $\Phi$ field, represented in red shades, in a fully developed flow scenario for the configuration $T_{jet}=15~^{\circ}\mathrm{C}$ and $u^{\prime}/U=20~\%$.

\begin{figure}[htb]
	\centering
	\begin{subfigure}[b]{0.8\columnwidth}
		\centering
		\includegraphics[width=\linewidth]{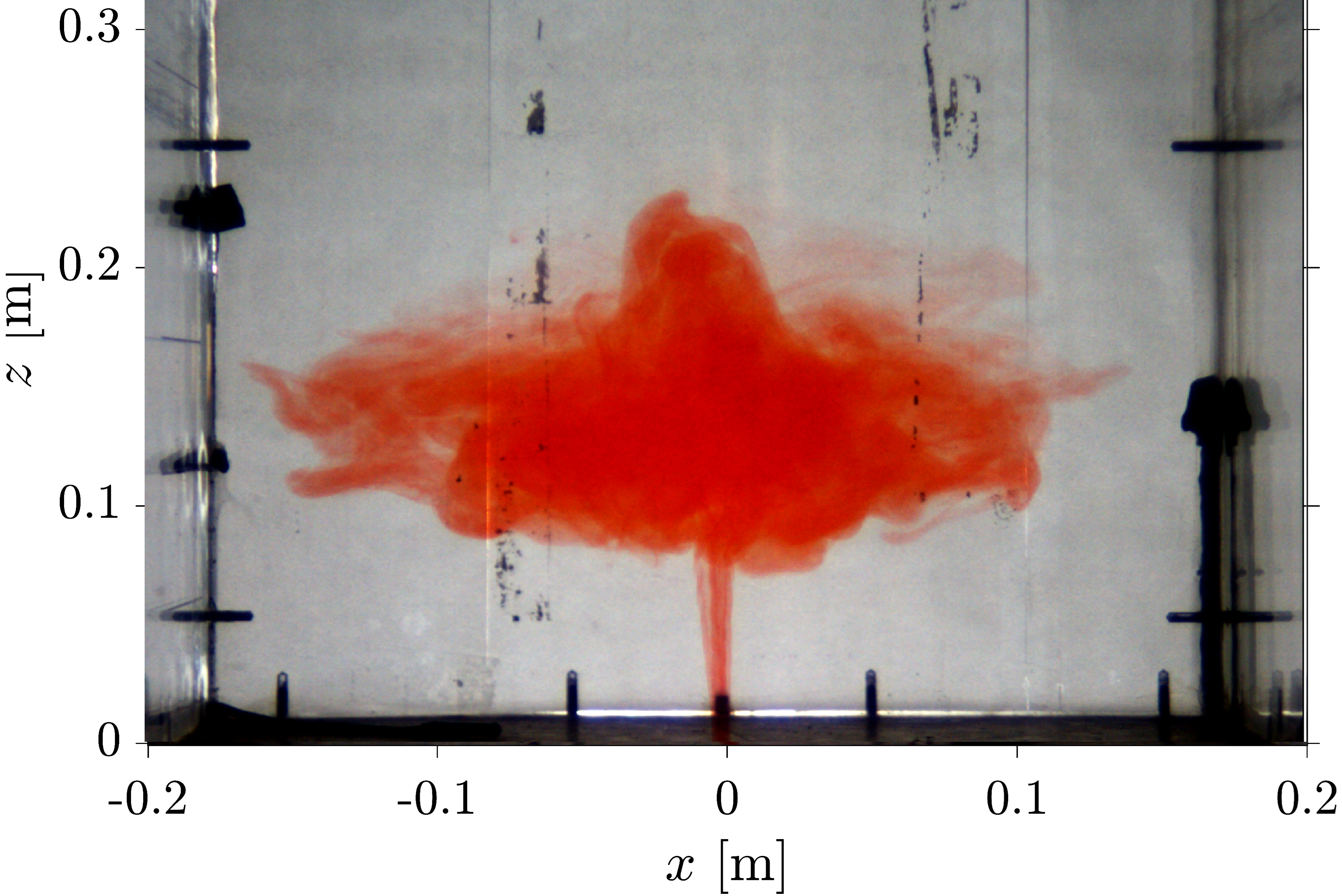}
		\caption{}
	\end{subfigure}
	%\hfill
	%\hspace{1pc}
	\begin{subfigure}[b]{0.8\columnwidth}
		\centering
        \includegraphics[width=\linewidth]{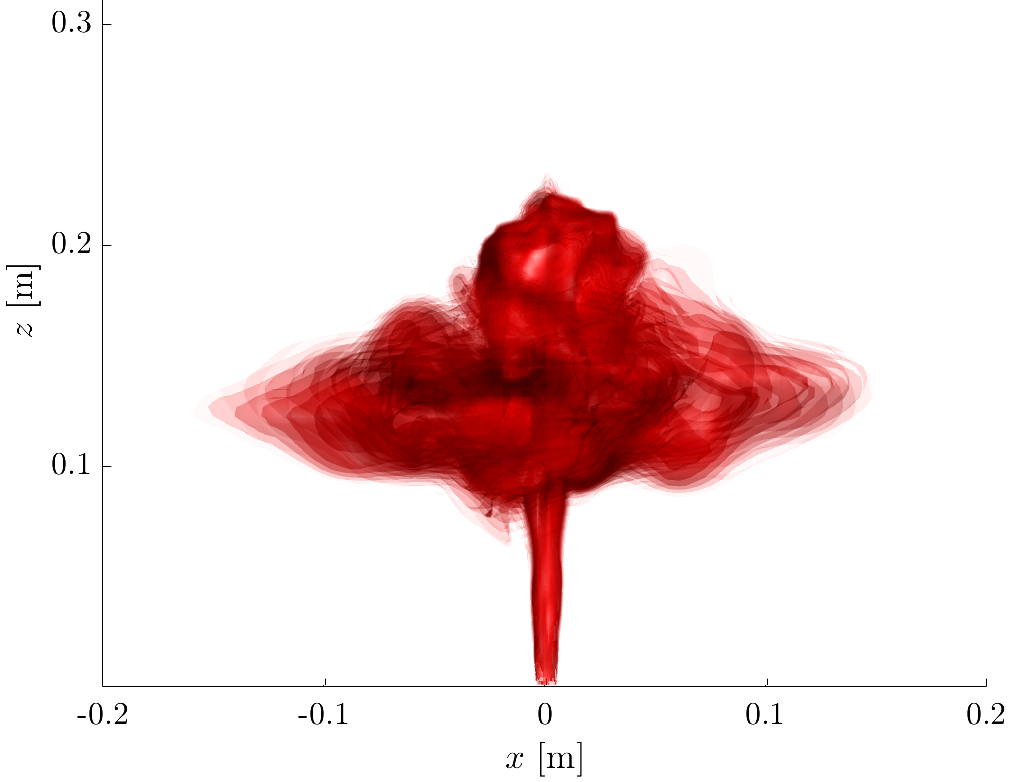}
		\caption{}
	\end{subfigure}
	\caption{\justifying Tracer fields of the fountain with fully developed flow at $t=80~\mathrm{s}$ for: (a) experimental measurements with the grid configuration and (b) computational results for a turbulence intensity of $u^\prime /U=20\%$.}
 \label{fig:campoTrazadorValidacion}
\end{figure}

The validation of the computational simulations was conducted by comparing the tracer fields with the previously performed experimental measurements~\cite{sarasua2021spreading, 2022FreireFTLE}. This was achieved by analysing both the shape of the fountain over time and the evolution of $h_m$ and $h_{sp}$. As an example, Fig.\ref{fig:campoTrazadorValidacion} compares a photograph of the tracer-marked fountain obtained experimentally for $T_{jet}=15~^{\circ}\mathrm{C}$, using an 80-wire-per-inch mesh screen (with each wire having a diameter of 0.18mm) placed transversely at the inlet to generate a specific turbulence level, referred to as the ``grid'' configuration, with the corresponding simulation for $u^{\prime}/U=20~\%$, both at $t=80~\mathrm{s}$. As can be seen, the overall shape of the fountain, as well as the heights it reaches, are in excellent agreement.

\begin{figure}[htb]
	\centering
	\includegraphics[width=0.8\columnwidth]{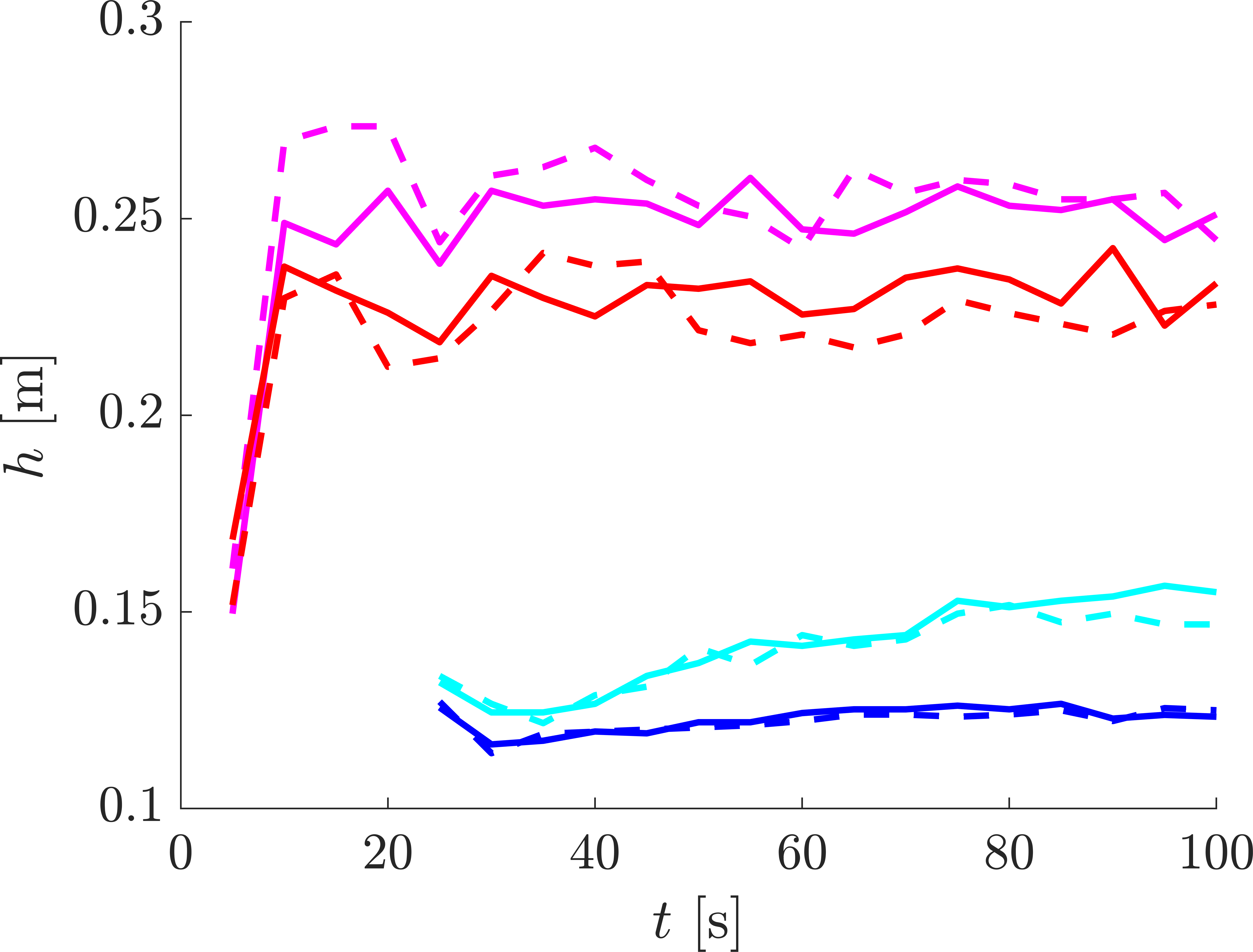}
	\caption{\justifying Evolution of $h_m$ and $h_{sp}$ obtained from previously conducted laboratory experiments~\cite{sarasua2021spreading,2022FreireFTLE} (solid lines) and computational simulations (dashed lines). In magenta (red), the evolution of $h_m$ for the free (grid) experimental configuration is shown, corresponding to a turbulence intensity of $u^{\prime}/U=2\%$ ($u^{\prime}/U=20\%$) in the computational simulations. In cyan (blue), $h_{sp}$ is shown for the free (grid) configuration. After $t\approx 70~\mathrm{s}$, all heights stabilise, and the flow becomes fully developed, as shown in Fig.~\ref{fig:campoTrazadorValidacion}.}
	\label{fig:alturasFuente}
\end{figure}

Additionally, Fig.~\ref{fig:alturasFuente} shows the evolution of $h_m$ and $h_{sp}$ for two experimental configurations: one referred to as ``free'', where $T_{jet}=15^{\circ}\mathrm{C}$ and no mesh screen was placed at the inlet, and the other being the grid configuration. These experimental results are shown alongside the computational simulation results for turbulence intensities of $u^{\prime}/U=2$ and $20~\%$, respectively. The computational and experimental data for the two configurations, free and grid, exhibit excellent agreement, further confirming the accuracy of the numerical simulations. This robust correlation between the simulations and laboratory experiments underscores the ability of the numerical model to faithfully reproduce the essential dynamics of turbulent fountains.
%This analysis demonstrates that the computational simulations are in excellent agreement with the experimental measurements, fully validating the numerical calculations.

%%%%%%%%%%%%%%%%%%%%%%%%%%%%

\section{\label{sec:indicators}Analysis and Results}

In this work, we analyse the impact of turbulent fountains on the lower layers of the ambient fluid as a function of two key parameters influencing the flow~\cite{sarasua2024influence}, namely the fountain characteristics, i.e., $T_{jet}$ and $u^{\prime}/U$. As in Sarasúa et al. (2024)~\cite{sarasua2024influence}, we work in the configuration space $\left(Fr^{-2},u^{\prime}/U\right)$, where $Fr$ represents the Froude number at the fountain inlet, which is defined as $Fr=U/\sqrt{g D \frac{\rho_{jet}-\rho_{00}}{\rho_{00}}}$, where $D$ is the fountain inlet diameter, $U$ is the mean inlet velocity, $\rho_{jet}$ is the density of the fountain at the inlet, and $\rho_{00}$ is the initial density of the stratified ambient fluid at $z=0$. These densities correspond to those of water at the given temperatures.

The performance of fountains with different characteristics is assessed using the efficiency indicators we propose, tailored to their specific applications. Firstly, their ability to mitigate low ambient fluid temperatures near ground level is evaluated in Sec.~\ref{subsec:temperature-increasing}. Secondly, in Sec.~\ref{subsec:tracer-detection}, we discuss the presence of a tracer as an indicator of the concentration of fluid that, after being ejected as a fountain, returns to the ground. Finally, in Sec.~\ref{subsec:protected-zone}, the radial threshold distance is determined, measured from the fountain’s point of emission, beyond which the tracer concentration falls below a specified tolerance value.

\subsection{\label{subsec:temperature-increasing}Average temperature change in the lower layers}% of the Ambient Fluid}

The phenomenon of entrainment (\textit{MTT} model~\cite{morton1956turbulent}) causes the fountain to incorporate fluid from its surroundings during its ascent, depending on the relative velocity conditions between the fountain and the ambient fluid, as well as the turbulence level of the ejected flow~\cite{sarasua2021spreading}. There is a delicate interaction of different mechanisms -advection, diffusion, and mixing- that governs both the dynamics of the fountain and the evolution of the ambient fluid around it. Previous literature focuses on analysing the characteristic heights reached by the fountain, i.e., the maximum, spreading, and minimal heights, but the evolution of the ambient fluid, especially near ground level where fountain action may have technological significance, has not been thoroughly addressed~\cite{guarga2000evaluation}.

We analyse the increase in the average ambient temperature in areas surrounding the fountain, within different radial distances $r$ from the centre of the fluid inlet, with $r\geqslant D$.
For this purpose, we first calculate the azimuthal average of the temperature field $T\left(x,y,z,t\right)$ to smooth out fluctuations inherent to the flow and obtain a temperature field as a function of the vertical coordinate $z$ and the radial distance from the fountain axis, $r$. Then, given the ambient fluid temperature at a point
%For this purpose, primero calculamos el promedo azimutal del campo de temperatures $T\left(x,y,z,t\right)$, para suavizar fluctuaciones propias del flujo y obtener así un campo de temperatures en función de la coordenada $z$ y de la distancia radial al eje de la fountain, $r$. Then, given the ambient fluid temperature at a point
$\left(r^{\prime},z^{\prime}\right)$ at time $t$, $T\left(r^{\prime},z^{\prime},t\right)$, and the initial temperature at the same point, $T\left(r^{\prime},z^{\prime},t=0\right) = T_0\left(z^{\prime}\right)$
%(since the initial stratification of the ambient fluid depends solely on height)
, we define the average temperature increase of the ambient fluid within the lower layers, below a specified height $z$, and at a radial distance less than $r$ (but greater than $D$), at time $t$, as
\begin{equation}
\Delta T_{\mathrm{avg}} \left(r,z,t\right) = 
\left<
T\left(r^{\prime},z^{\prime},t\right) - T_0\left(z^{\prime}\right)
\right>_{\left(r^{\prime},z^{\prime}\right)\in \mathfrak{R}\left(r,z\right)}
\label{eq:defDeltaT}
\end{equation}
where $\left< f \right>_{\mathfrak{R}\left(r,z\right)}$ represents the spatial average of a scalar field $f$ over the region $\mathfrak{R}\left(r,z\right)$, defined as
\begin{equation}
\mathfrak{R}\left(r,z\right) = 
\left\{ \left(r^{\prime},z^{\prime}\right) \mathrm{, such~that~}
D\leqslant r^{\prime} \leqslant r \mathrm{~and~} 0\leqslant z^{\prime} \leqslant z \right\}
\label{eq:defRegionR}
\end{equation}
Here, $\Delta T_{\mathrm{avg}}$ is expressed in degrees Celsius ($^{\circ}\mathrm{C}$), although it can equally be expressed in Kelvin ($\mathrm{K}$), as it represents a temperature difference.

\begin{figure}[htb]
	\centering
	\begin{subfigure}[b]{0.8\columnwidth}
		\centering
		\includegraphics[width=\linewidth]{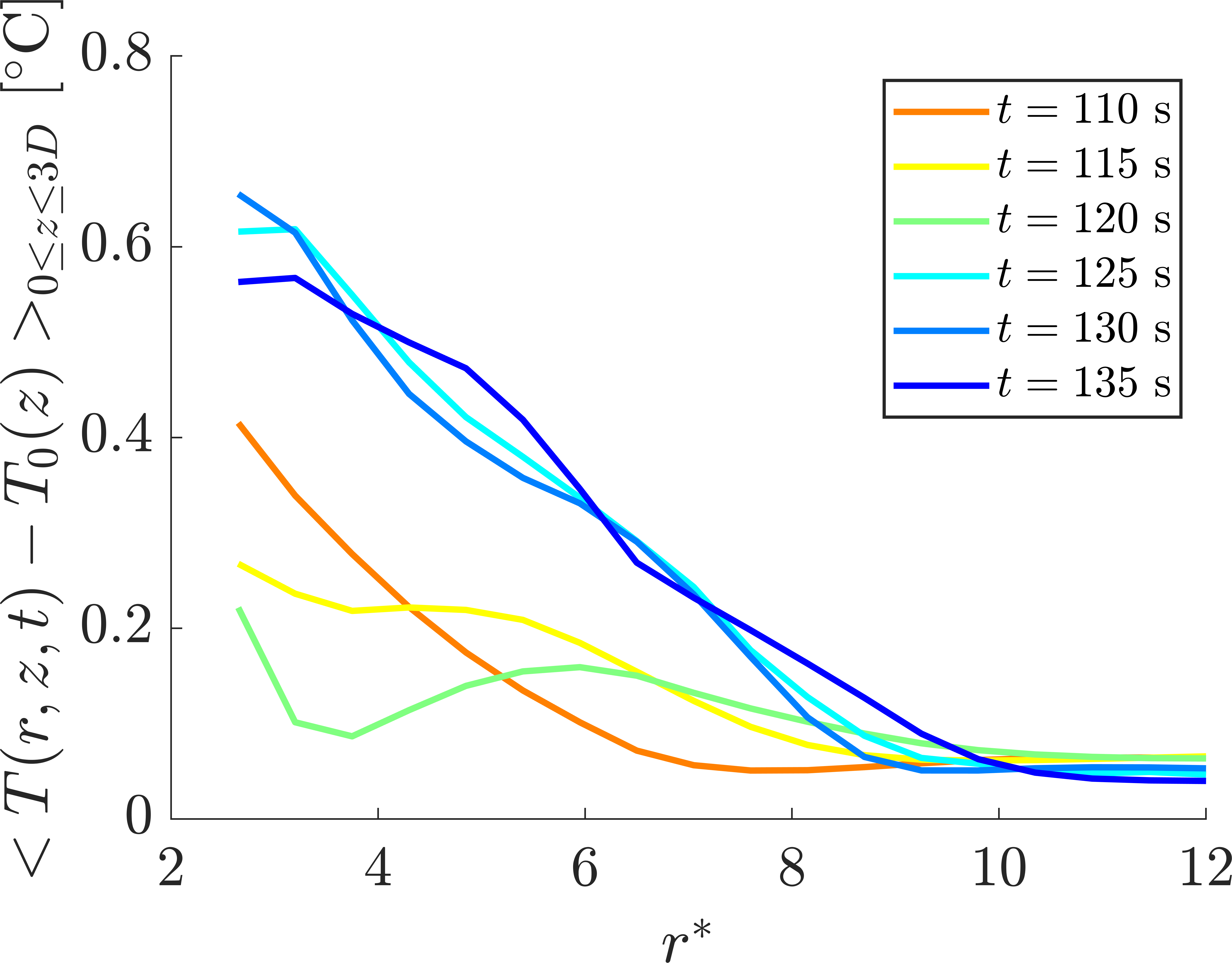}
		\caption{}
	\end{subfigure}
	%\hfill
	%\hspace{1pc}
	\begin{subfigure}[b]{0.8\columnwidth}
		\centering
		\includegraphics[width=\linewidth]{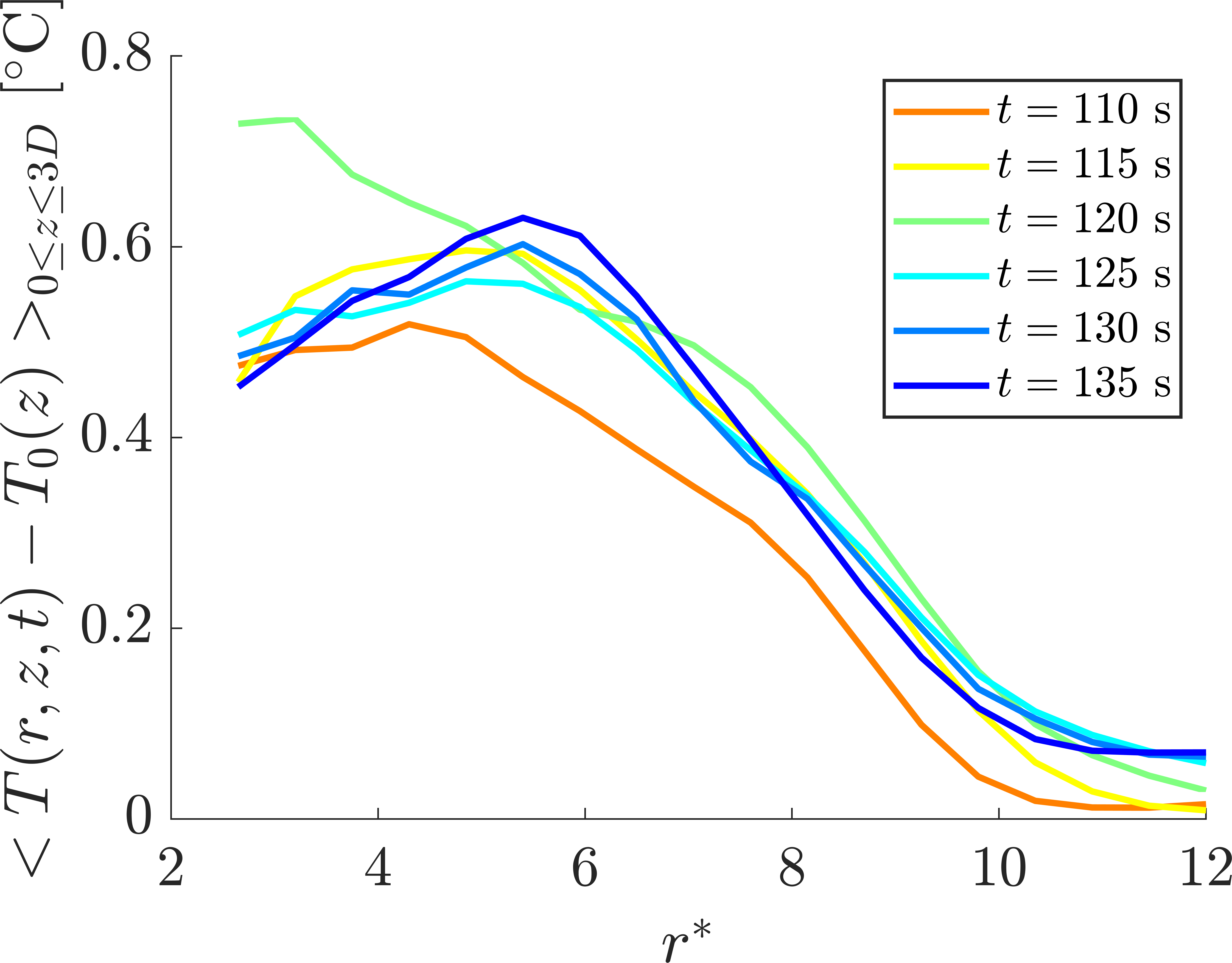}
		\caption{}
	\end{subfigure}
	\caption{\justifying
    %\textcolor{red}{Dejo la pregunta planteada, no para este trabajo si no para más adelante: se podrá adimensionar estos perfiles (y/o los de fig8) y ver si colapsan para $t>\tau$? (Nica)}
    Profiles of $\left< T\left(r,z,t\right) - T_0\left(z\right) \right>_{0\leqslant z \leqslant 3D}$ for different time instants as a function of $r^\ast$, for the fountain configurations: (a) $(Fr^{-2},u^\prime /U)=(3.8\times10^{-3}, 4\%)$, and (b) $(Fr^{-2},u^\prime /U)=(4.9\times10^{-3}, 2\%)$.}
 \label{fig:perfiles_promedioAumentoTemperatura}
\end{figure}

For this analysis, we first determine the moment from which the conditions in the lower layers of the ambient fluid stabilise. Here, we arbitrarily consider the lower layers of the ambient fluid as those located in $0\leqslant z \leqslant 3D$. We then seek to determine the time $\tau$ such that the shape of the profile $\left< T\left(r,z,t\right) - T_0\left(z\right) \right>_{0\leqslant z \leqslant 3D}$ as a function of $r$ stabilises for $t\geqslant \tau$. In Fig.~\ref{fig:perfiles_promedioAumentoTemperatura}, this profile is shown at different time instants for two arbitrarily chosen fountain configurations as a function of the non-dimensional radial distance $r^{\ast}=r/D$ (with $r^{\ast}=25$ being the maximum radius of the domain).
As observed, the profiles cease to vary (except for inherent turbulent flow fluctuations) and stabilise into a steady profile in the fully developed flow regime at $\tau=125$~s. Although this analysis is presented here for two arbitrary configurations, this stabilisation time was explored for the other configurations considered in this work, and the same behaviour was consistently observed.
%\sout{This convergence time is consistent across all computational configurations resolved in this work.}\textcolor{red}{2 comentarios:1- capaz que para cubrirnos pondría que los perfiles convergen con una diferencia menor al XX\% (o algún criterio del estilo) porque en la fig 6b a ojo no me resulta tan evidente. 2- (Vinculado al comentario de las configuraciones) Acá dice todas las configuraciones y en la fig 6 muestra dos configs, hay alguna más ? (Nica)}

\begin{figure}[htb]
	\centering
	\begin{subfigure}[b]{0.8\columnwidth}
		\centering
		\includegraphics[width=\linewidth]{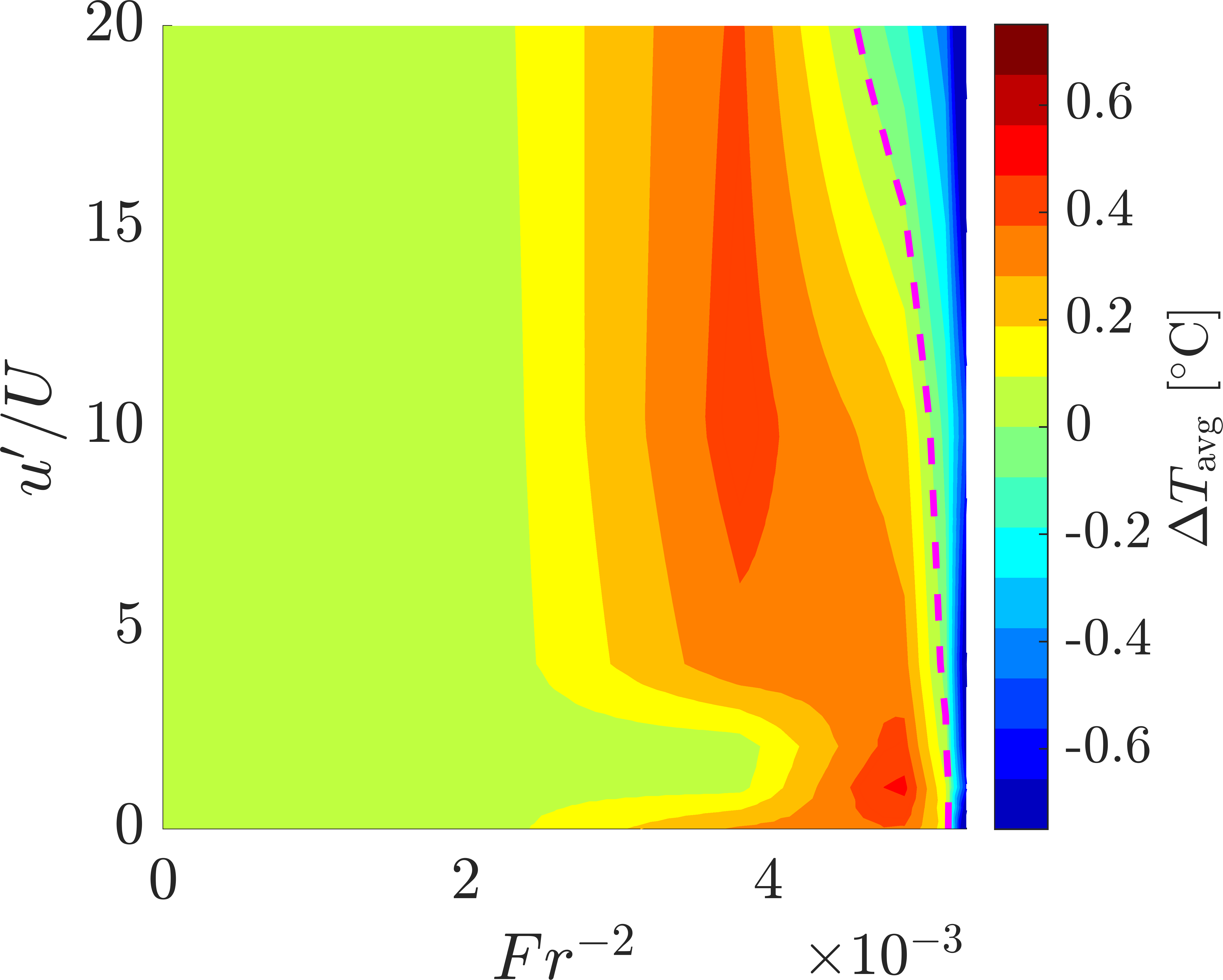}
		\caption{}
	\end{subfigure}
	%\hfill
	%\hspace{1pc}
	\begin{subfigure}[b]{0.8\columnwidth}
		\centering
		\includegraphics[width=\linewidth]{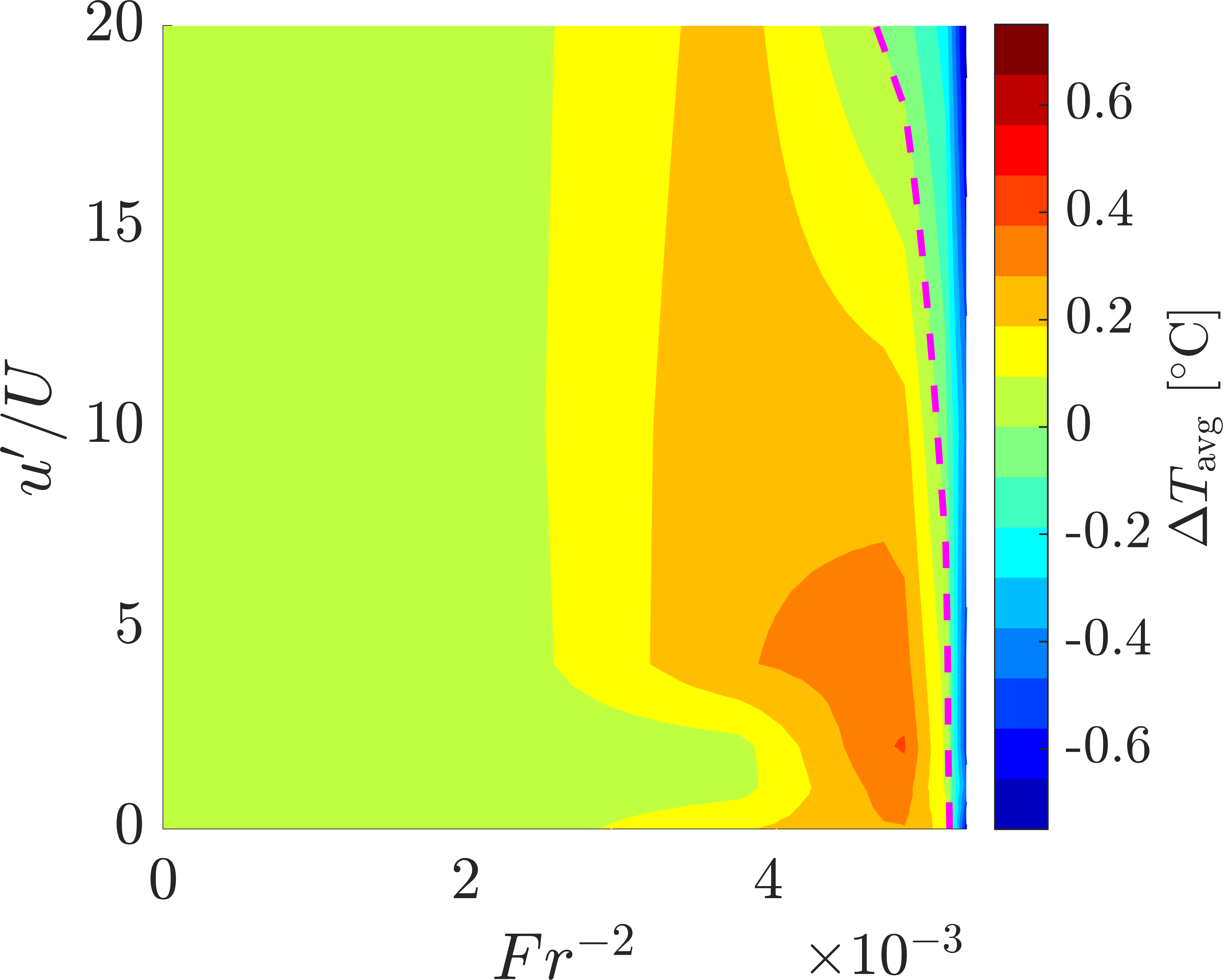}
		\caption{}
	\end{subfigure}
	\caption{\justifying Two-dimensional colour map of the mean temperature increase in the lower layers of the ambient fluid: (a) $\Delta T_{\mathrm{avg}} \left(r=5D,~z=3D,~t=135s\right)$, and (b) $\Delta T_{\mathrm{avg}} \left(r=10D,~z=3D,~t=135s\right)$. The magenta dashed line marks the right boundary of the configuration space where $\Delta T_{\mathrm{avg}}$ is positive.}
	\label{fig:promedioAumentoTemperatura}
\end{figure}

Based on the time at which the temperature increase profiles converge, we calculate the net ambient temperature rise, $\Delta T_{\mathrm{avg}}$, from Eq.~\ref{eq:defDeltaT}, for the different computational configurations at time $t=135$s. In Fig.\ref{fig:promedioAumentoTemperatura}(a), we present a colour map of $\Delta T_{\mathrm{avg}} \left(r=5D,~z=3D,~t=135s\right)$ in the configuration space $(Fr^{-2}, u^\prime /U)$. Fig.\ref{fig:promedioAumentoTemperatura}(b) shows the corresponding data for $\Delta T_{\mathrm{avg}} \left(r=10D,~z=3D,~t=135s\right)$. The general pattern of the two maps is similar, indicating that the average increase in ambient fluid temperature in the lower layers is largely independent of radial distance.
The impact of the flow on the near-ground ambient is particularly noteworthy. Firstly, for $Fr^{-2}~\leqslant~5\times10^{-3}$, i.e., fountain temperatures above approximately 4~$^{\circ}\mathrm{C}$, there is always an increase in the average temperature of the lowest layers of the ambient fluid. The largest temperature increase is observed primarily in a region of the configuration space with $2.2\times10^{-3}~\leqslant~Fr^{-2}~\leqslant~5.3\times10^{-3}$ (fountain temperatures in the range of approximately 4–12.5~$^{\circ}\mathrm{C}$). However, this region is irregular in shape and narrows for turbulence intensities in the range $1\% \geqslant u^{\prime}/U \geqslant 3\%$, covering the interval $3.8\times10^{-3}\leqslant Fr^{-2} \leqslant 5.3\times10^{-3}$ (fountain temperatures in the range of approximately 4–10~$^{\circ}\mathrm{C}$).

This behaviour is noteworthy but can be explained by the fountain collapse diagram from Sarasúa et al. (2024)\cite{sarasua2024influence}, which outlines the regions of the configuration space $\left( Fr^{-2} , u^{\prime}/U \right)$ corresponding to non-collapse, semi-collapse, and full collapse of the fountain. The collapse region, i.e., configurations where the spreading height is $h_{sp}=0$, is indicated for $Fr^{-2} \geqslant 5.3\times10^{-3}$, regardless of turbulence intensity. In this region, a cooling of the lower layers of the ambient fluid is observed, as shown in Fig.~\ref{fig:perfiles_promedioAumentoTemperatura}. Given that, during fountain collapse, fluid with a high tracer concentration is detected at ground level, meaning it has undergone little mixing with the ambient fluid during its ascent, and that this mixing only occurs up to small maximum fountain heights ($h_m$), the temperature of the fountain fluid returning to the ground is low enough to cause the cooling of the lower ambient fluid layers.

On the other hand, for $Fr^{-2} \leqslant 5.3\times10^{-3}$, the collapse diagram separates the regions of non-collapse and semi-collapse, both characterised by $h_{sp}>0$. To distinguish between these regimes, an arbitrary tracer concentration tolerance, $\Phi_{tol}$, is introduced in Sarasúa et al. (2024)~\cite{sarasua2024influence}. The non-collapse regime is defined as the set of configurations where, once the fully developed flow state is reached, $\Phi\left(x,~y,~z,~t\right)<\Phi_{tol}$ across \textit{the entire ground level}, i.e., $z=0$, while the semi-collapse region corresponds to configurations where $\Phi>\Phi_{tol}$ \textit{at any point} in the ambient fluid at ground level. It is important to note that, for constructing the collapse diagram in such work, $\Phi$ was considered within the ambient fluid layer near the ground, rather than averaged over the lower layers.

The region of the configuration space where the greatest heating of the lower layers of the ambient fluid is detected (see Fig~\ref{fig:promedioAumentoTemperatura}) lies within the semi-collapse region of the collapse diagram, regardless of the value of $\Phi_{tol}$ considered. Therefore, the semi-collapse of the fountain promotes the average increase in temperature of the lower layers of the ambient fluid. To better understand the underlying mechanism of this phenomenon, it is necessary to reference the minimal height, $h_c$, defined in Sarasúa et al. (2024)\cite{sarasua2024influence} as the lower boundary of the spreading cloud (which is outlined by the tracer in the fully developed flow regime, where the ejected fountain fluid spreads radially, moving away from the fountain; see Fig.~\ref{fig:esuemaFuente}). In the configurations where heating of the lower layers of the ambient fluid is detected in Fig.~\ref{fig:promedioAumentoTemperatura}, we observe that $h_c>0$ and it increases as $Fr^{-2}$ decreases. Thus, it is easy to envision how, during the transient stage of the flow and up until the fully developed flow regime, the spreading cloud displaces ambient fluid from higher layers, but below $h_{sp}$, downwards (such motion is represented with red arrows in Fig.~\ref{fig:esuemaFuente}). This fluid mixes with the lower layers of the ambient fluid, leading to more efficient heating of these layers.

Supporting this potential heating mechanism, the collapse diagram shows a narrowing of the semi-collapse region for lower turbulence intensities, which coincides with the narrowing of the region of greatest heating in the lower layers of the ambient fluid, as explained by the results in Fig.~\ref{fig:promedioAumentoTemperatura}. From this analysis, we observe that the evolution of the ambient fluid is strongly dependent on the characteristics of the flow induced by the fountain during the fully developed flow stages.

\subsection{\label{subsec:tracer-detection}
%Volume of Ejected Fluid Returning to the Ground: Average Tracer Concentration in the Lower Layers of Ambient Fluid
Return of ejected fluid:  Tracer concentration in the lower layers
}

Let us assume, motivated by the applications of the SIS device introduced in Sec.~\ref{sec:intro}, that the fountain fluid, which is denser than the ambient fluid, represents a fluid containing some type of contaminant. It is crucial, therefore, to determine how much ``contaminated'' fluid, depending on the fountain configuration, comes back into contact with the ground.
In this work, we perform a detailed analysis of the average tracer concentration in the lower layers of the ambient fluid within two radial distances from the fountain entry.

The proposed method first involves, similar to Sec.~\ref{subsec:temperature-increasing}, obtaining the field $\phi(r,z,t)$, which represents the azimuthal average of the tracer field $\Phi\left(x,~y,~z,t\right)$ around the fountain axis, $r=0$, at each time instant. Then, analogous to the procedure applied to $\Delta T_{\mathrm{avg}} \left(r,z,t\right)$ in Eq.~\ref{eq:defDeltaT}, we define the average tracer concentration over the lower layers of the ambient fluid at time $t$, $\phi_{\mathrm{avg}} \left(r,~z,~t\right)$, as
\begin{equation}
\phi_{\mathrm{avg}} \left(r,z,t\right) = 
\left<
\phi\left(r^{\prime},z^{\prime},t\right)
\right>_{\left(r^{\prime},z^{\prime}\right)\in \mathfrak{R}\left(r,z\right)}
\label{eq:defPromTracer}
\end{equation}
where $\mathfrak{R}\left(r,z\right)$ is the region defined in Eq.~\ref{eq:defRegionR}.

\begin{figure}[htb]
	\centering
	\begin{subfigure}[b]{0.8\columnwidth}
		\centering
		\includegraphics[width=\linewidth]{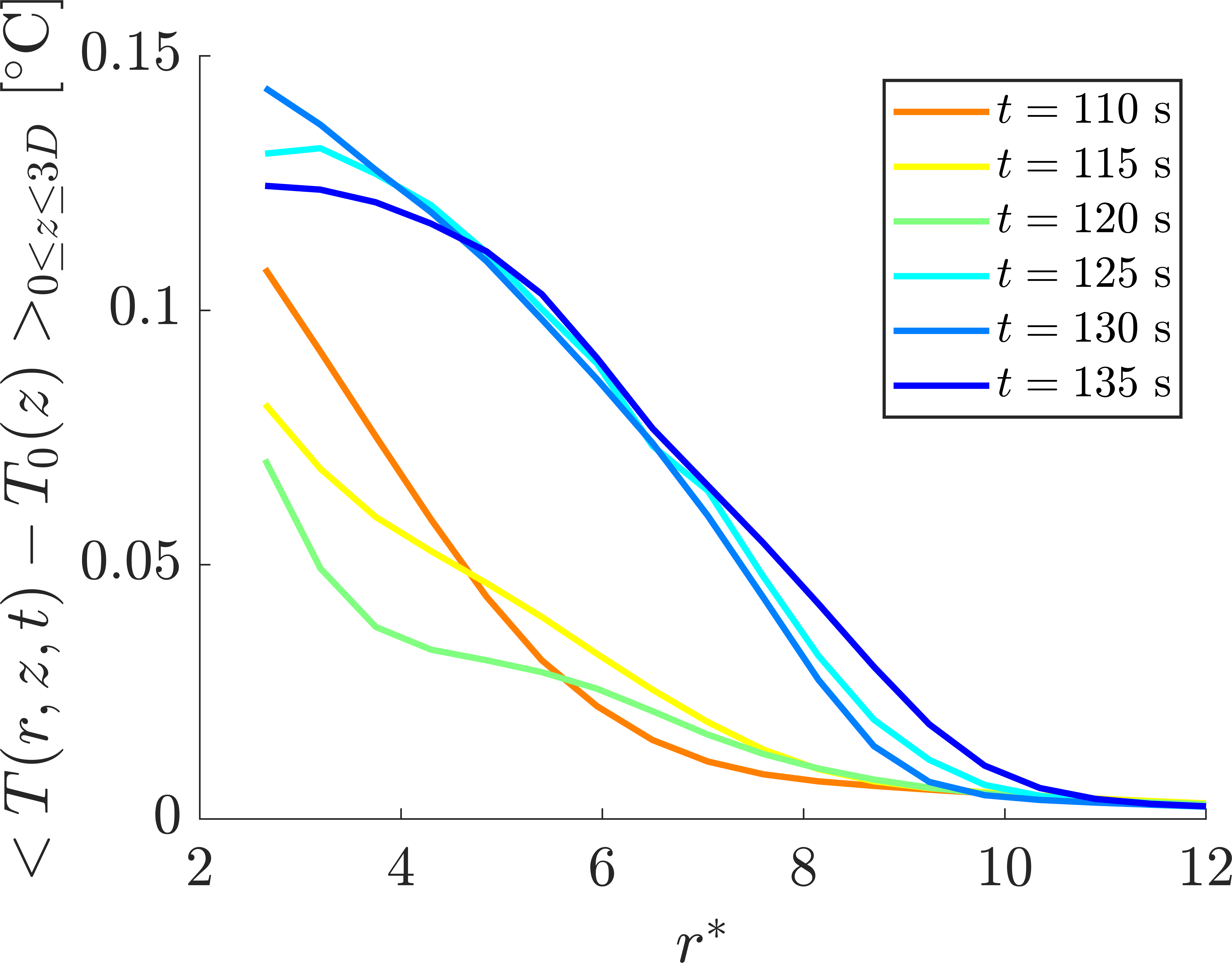}
		\caption{}
	\end{subfigure}
	%\hfill
	%\hspace{1pc}
	\begin{subfigure}[b]{0.8\columnwidth}
		\centering
		\includegraphics[width=\linewidth]{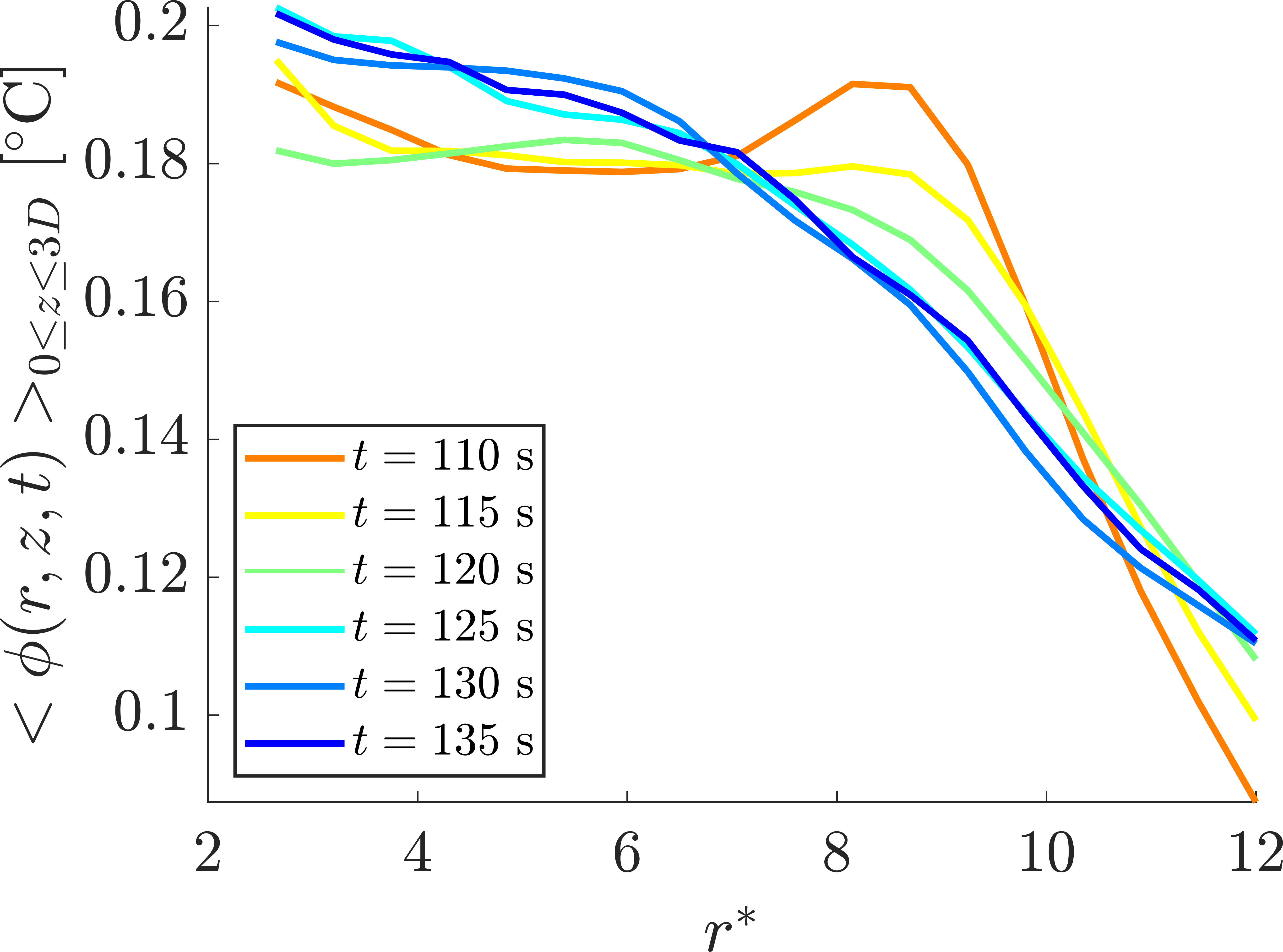}
		\caption{}
	\end{subfigure}
	\caption{\justifying Profiles of $\left< \phi\left(r,z,t\right)\right>_{0\leqslant z \leqslant 3D}$ as a function of $r^{\ast}$, for different time instants, and the configurations: (a) $(Fr^{-2},u^\prime /U)=(3.8\times10^{-3}, 4\%)$, and (b) $(Fr^{-2},u^\prime /U)=(4.9\times10^{-3}, 2\%)$.}
	\label{fig:perfiles_promedioTrazador}
\end{figure}

To determine the moment $\tau$ from which the average tracer concentration in the lower layers of the ambient fluid stabilises, we study, similarly to what was done in Sec.\ref{subsec:temperature-increasing}, the evolution of the profile $\left< \phi\left(r,z,t\right)\right>_{0\leqslant z \leqslant 3D}$ at different time instants. Figure~\ref{fig:perfiles_promedioTrazador} shows such profiles for two arbitrarily selected fountain configurations. In all cases, we observe that for $t\geqslant 125$~s, the average tracer concentration ceases to evolve. Therefore, in this section, we choose the time $t=135$s, as we did with the average temperature increase in Sec.~\ref{subsec:temperature-increasing}, to analyse the average tracer field.

\begin{figure}[htb]
	\centering
	\begin{subfigure}[b]{0.8\columnwidth}
		\centering
		\includegraphics[width=\linewidth]{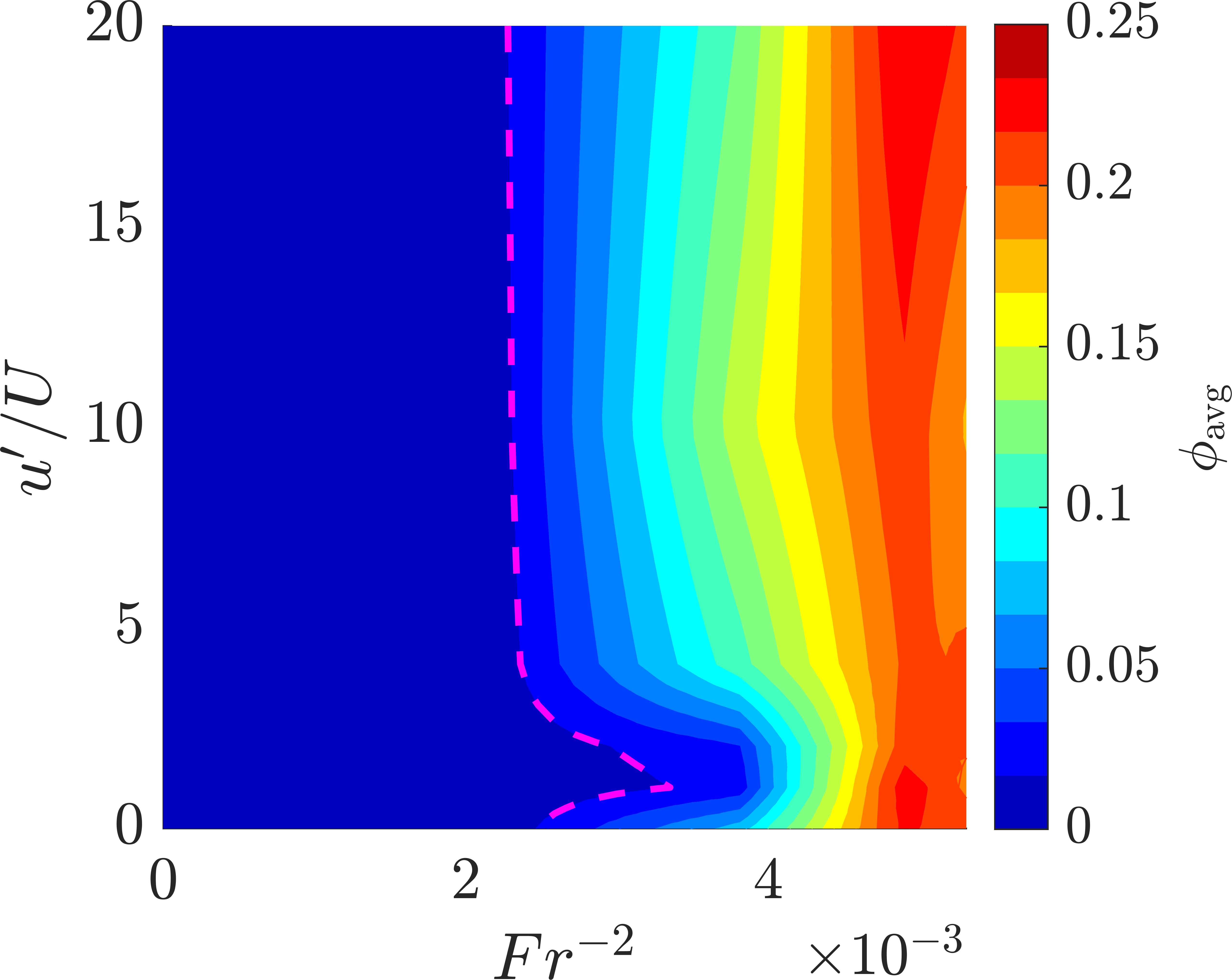}
		\caption{}
	\end{subfigure}
	%\hfill
	%\hspace{1pc}
	\begin{subfigure}[b]{0.8\columnwidth}
		\centering
		\includegraphics[width=\linewidth]{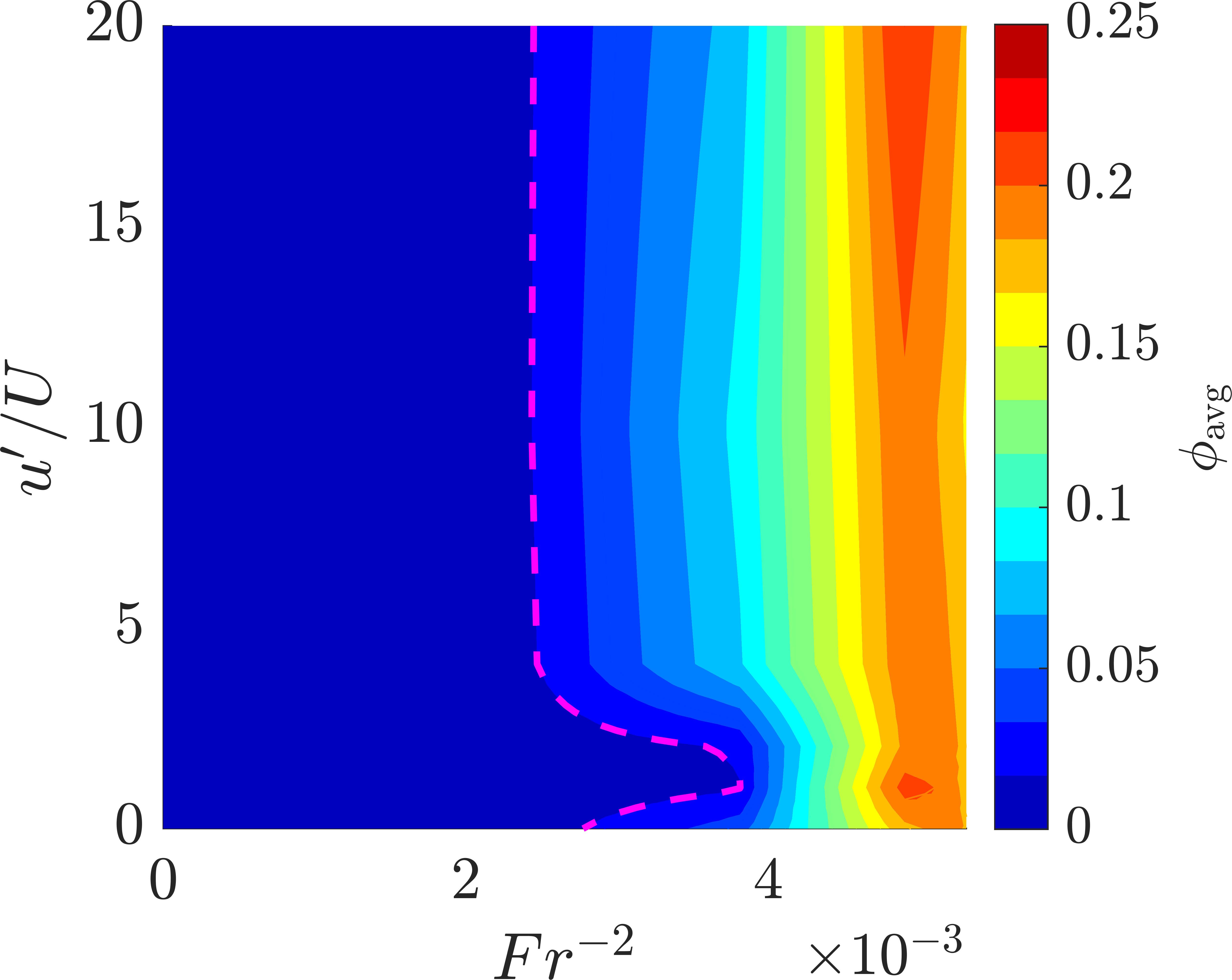}
		\caption{}
	\end{subfigure}
	\caption{\justifying Two-dimensional colour map of the mean tracer concentration in the lower layers of the ambient fluid: (a) $\phi_{\mathrm{avg}}\left(r=5D,~z=3D,~t=135\mathrm{s}\right)$, and (b) $\phi_{\mathrm{avg}}\left(r=10D,~z=3D,~t=135\mathrm{s}\right)$. The magenta dashed line represents the contour where $\phi_{\mathrm{avg}}$ equals 0.01.}
	\label{fig:promedioTrazador}
\end{figure}

Once the time instant for analysis is determined, in Fig.~\ref{fig:promedioTrazador}(a) we show the colour map of $\phi_{\mathrm{avg}}\left(r=5D,~z=3D,~t=135\mathrm{s}\right)$, and Fig.\ref{fig:promedioTrazador}(b) shows the corresponding map for $\phi_{\mathrm{avg}}\left(r=10D,~z=3D,~t=135\mathrm{s}\right)$. The first observation from these figures is, as in Sec.~\ref{subsec:temperature-increasing}, their strong similarity, indicating that when the fountain fluid comes into contact with the lower layers of the ambient fluid, the tracer distribution is relatively uniform in the radial direction for $r^{\ast} \leqslant 10$. 

Secondly, by comparing Figures~\ref{fig:promedioAumentoTemperatura} and \ref{fig:promedioTrazador}, we observe an overlap between the regions of temperature increase in the ambient fluid and the detection of the tracer in the lower layers. This overlap indicates that the presence of fountain fluid returning to the ground after mixing with the warmer, upper layers of ambient fluid contributes to the overall increase in the mean temperature of the lower layers.

Furthermore, let us consider a region in the configuration space where the temperature increase is most pronounced. As seen in Fig.~\ref{fig:promedioAumentoTemperatura}, this occurs, for example, at $Fr^{-2}=3.8\times10^{-3}$ and $u^{\prime}/U > 6\%$. If we examine the average tracer concentration in the lower layers of the ambient fluid for the same region, as shown in Fig.~\ref{fig:promedioTrazador}, the values range between $0.10< \phi_{\mathrm{avg}} < 0.15$. This suggests a low concentration of fountain fluid, indicating that most of the fluid present is ambient fluid that was entrained and displaced from the upper layers of the environment. This finding reinforces the hypotheses presented in Sec.~\ref{subsec:temperature-increasing} regarding the mechanisms that govern the conditions of the lower layers of the ambient fluid.
As we move away from this configuration of maximum ambient heating, two scenarios arise: either we shift to configurations with lower $Fr^{-2}$ (i.e., higher $T_{jet}$), where the fountain reaches greater heights and does not displace significant amounts of warmer ambient fluid downwards, or we shift to configurations with higher $Fr^{-2}$ (i.e., lower $T_{jet}$), where the fountain reaches lower heights and tends to settle near the ground, with minimal mixing with the warmer upper layers of the ambient fluid, and thus without having significantly increased its temperature beforehand.

By identifying and supporting this correlation between the average tracer concentration and the average temperature increase in the lower layers of ambient fluid, we can explain the behaviour observed at low turbulence intensities, between $1~\%$ and $3~\%$, as shown in Fig.~\ref{fig:promedioTrazador}. In these cases, the fountain configurations that result in zero tracer concentration in the lower layers extend to higher values of $Fr^{-2}$ compared to those with higher or lower turbulence intensities.

To conclude our study, Sec.~\ref{subsec:protected-zone} examines the size of the region affected by the return of the fountain fluid, with the aim of predicting the real impact of the fluid that returns to the ground, beyond the average concentration detected in the lower layers of the ambient fluid.

\subsection{\label{subsec:protected-zone} Pollutant-free locations}

\begin{figure}[htb]
	\centering
	\begin{subfigure}[b]{0.8\columnwidth}
		\centering
		\includegraphics[width=\linewidth]{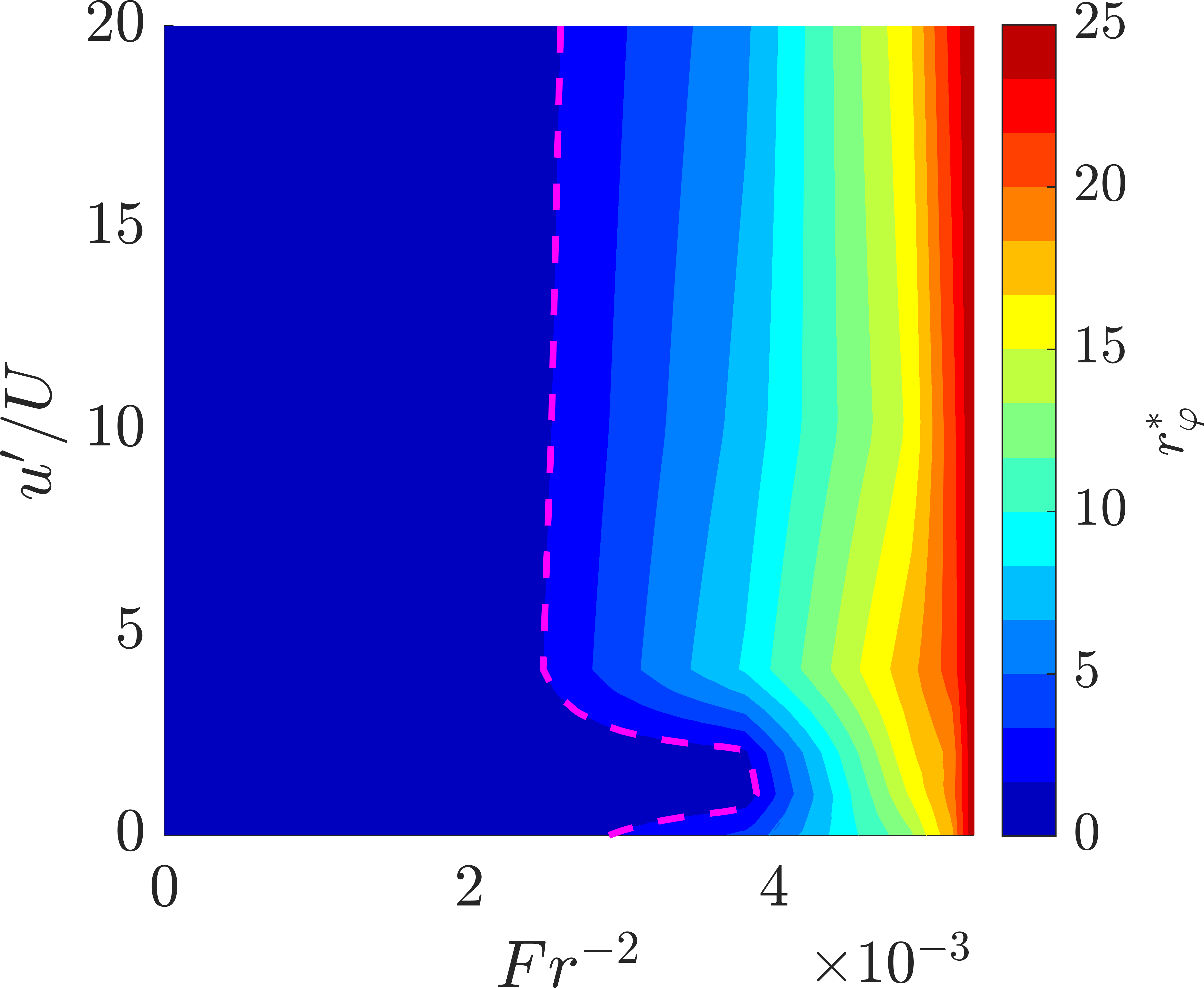}
		\caption{}
	\end{subfigure}
	%\hfill
	%\hspace{1pc}
	\begin{subfigure}[b]{0.8\columnwidth}
		\centering
		\includegraphics[width=\linewidth]{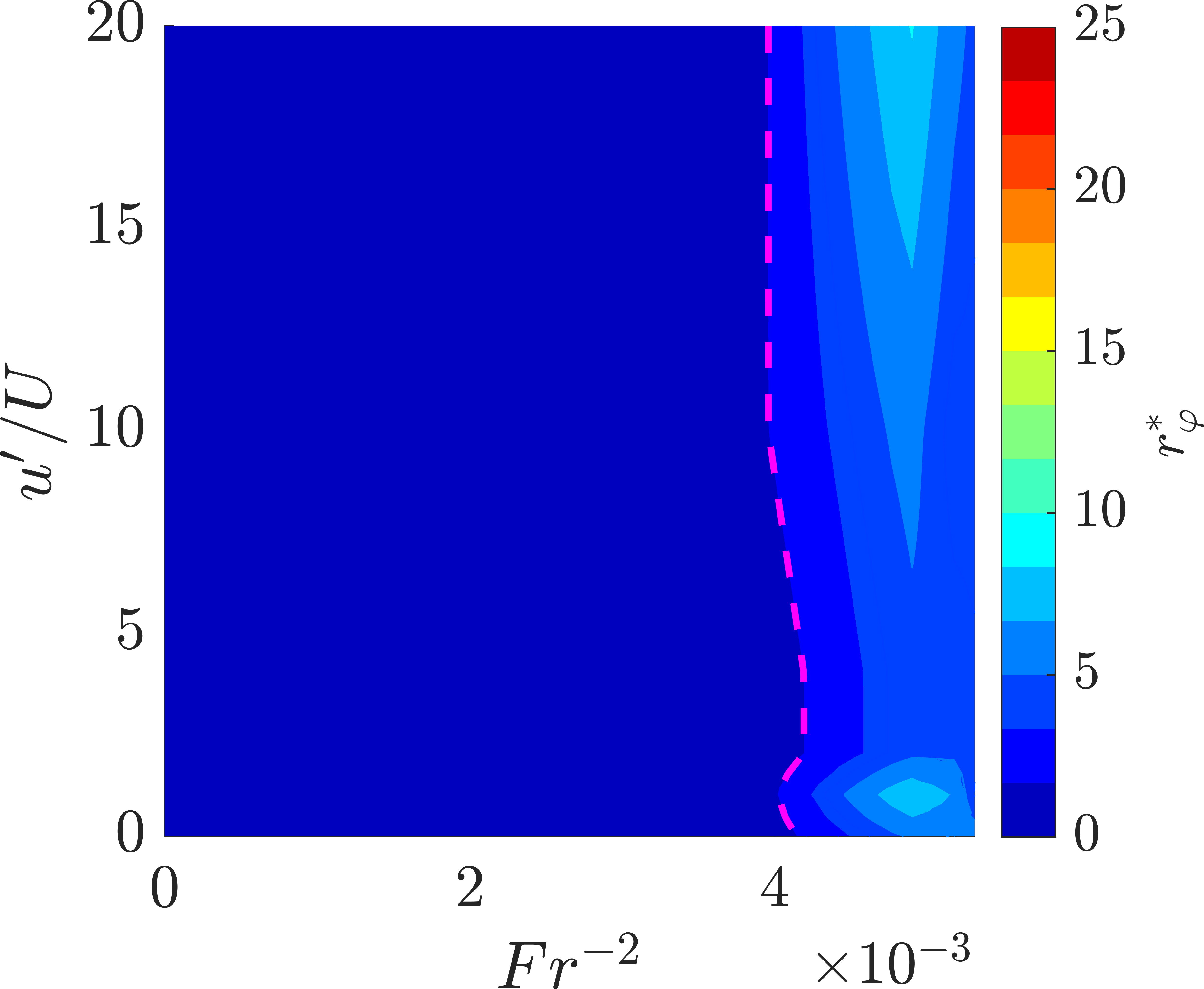}
		\caption{}
	\end{subfigure}
	\caption{\justifying Two-dimensional colour map of the radial distance $r^\ast$ at which a mean tracer concentration $r_\varphi\left(z=3D,t=135\mathrm{s}\right)$ is detected in the lower layers of the ambient, with: (a) $\varphi = 0.05$ and (b) $\varphi =  0.20$. The magenta dashed line marks the right boundary of the configuration space free of tracer, i.e., where the concentration is below the given threshold $\varphi$.}
	\label{fig:radioPorTrazador}
\end{figure}

%NUEVO
In this section, we delve deeper into the analysis of the impact of the ejected fluid from the fountain that returns to the ground, investigating which areas are affected. For a given tracer concentration $\varphi$, we define the impact radius $r_\varphi$ as follows:
\begin{equation}
\displaystyle
r_\varphi\left(z,t\right)=\max \left\{
r\geqslant D\mathrm{,~such~that~}
\phi_{\mathrm{avg}} \left(r,z,t\right) \geqslant \varphi
\right\}
\label{eq:radio_de_impacto}
\end{equation}
Figure~\ref{fig:radioPorTrazador} shows colour maps of the non-dimensional impact radius, i.e., $r^{\ast}_\varphi=r_\varphi /D$, for the same layers of ambient fluid and the same time instant chosen in the previous sections, i.e., $r_\varphi\left(z=3D,t=135\mathrm{s}\right)$, in the configuration space $\displaystyle \left(Fr^{-2}, u^{\prime}/U\right)$, for two values of $\varphi$: (a) $\varphi=0.05$ and (b) $\varphi=0.20$. As expected, $r^{\ast}_\varphi$ decreases as $\varphi$ increases for each fountain configuration. Specifically, for a tracer threshold of $0.20$, the ground remains unaffected for $Fr^{-2}\leqslant 4\times10^{-3}$, i.e., fountain temperatures below 10~$^{\circ}\mathrm{C}$, regardless of inlet turbulence intensity.

%NUEVO
For low turbulence intensities in the range of approximately $1~\%$ to $3~\%$, Fig.~\ref{fig:radioPorTrazador}(a) shows that the lower layers of the ambient fluid remain unaffected by the return of ejected fluid over a broader range of $Fr^{-2}$, up to approximately $3.8\times10^{-3}$. For higher fluctuations, these ambient layers are unaffected by the fountain return fluid for $Fr^{-2}\geqslant 2.2\times10^{-3}$, i.e., fountain temperatures above 12.5~$^\circ \mathrm{C}$. This is in excellent agreement with the analyses in Secs.~\ref{subsec:temperature-increasing} and \ref{subsec:tracer-detection}, where we observed similar irregularities in regions of increased ambient fluid heating and lower tracer concentrations, respectively.

By defining the impact radius in this manner, we can assess the extent of the regions affected by the return of the fountain fluid for various levels of turbulence and stratification. This metric provides a practical way to quantify the influence of fountains on the ambient environment, allowing for more precise predictions of the areas impacted under different configurations.

\section{Final remarks} \label{sec:conc}

Here, we provided tools to predict the impact of a fountain injected into a linearly stratified environment on the lower layers of the atmosphere.
The evolution of the ambient fluid exhibits distinct behaviour at low turbulence intensities. In this regime, a lower inlet temperature of the fountain, i.e., higher $Fr^{-2}$, hinders the increase in the average temperature of the lower ambient fluid layers. However, it simultaneously increases the protected distance from the return of ejected fluid, i.e., the impact radius is extended.

The most notable behaviour is observed in the effect on the temperature of the lower ambient layers. For $Fr^{-2}\leqslant 5.3\times10^{-3}$, i.e., injected fountain temperatures above 4~$^\circ \mathrm{C}$, the temperature of the lower ambient layers consistently increases. However, there exists an optimal region in the configuration space $\left(Fr^{-2},u^{\prime}/U\right)$, where the temperature increase is maximised. This region is independent of the fountain’s turbulence intensity for $2.2\times10^{-3}\leqslant Fr^{-2} \leqslant 5.3\times10^{-3}$, except for $1~\% \leqslant u^{\prime}/U \leqslant 3~\%$, where the region narrows, and the temperature increase occurs for $Fr^{-2} \geqslant 3.8\times10^{-3}$, i.e., fountain temperatures below 10~$^\circ \mathrm{C}$. This behaviour can be explained based on the work of Sarasúa et al. (2024)~\cite{sarasua2024influence}, as this region coincides with the semi-collapse regime of the fountain. In this regime, part of the fountain fluid returns to the surface but at a lower tracer concentration, indicating that it has already mixed with ambient fluid at a higher level, i.e., at a higher temperature, which deposits this warmer fluid in the lower layers, thereby increasing the ambient temperature. The temperature evolution is thus a delicate balance between the fountain collapse regime and the characteristic heights reached by the fountain, governed by its inlet properties.

Although this model is based on conditions typically observed in the lower atmosphere during radiation frosts, the analysis is also applicable to other stratified conditions in ambient fluids.
As future work, a highly interesting analysis from a technological perspective would be to study the case of injecting a plume, where $T_{jet}$ is higher than the ambient fluid temperature at ground level. After rising to a certain height, the plume will encounter a warmer layer of fluid and will transform into a fountain. This analysis would provide insight into the potential benefits of heating the fluid ejected by the SIS device.

%%%%%%%%%%%%%%%%%%%%%%%%%%%%%

\section*{\label{sec:ack} Acknowledgements}

The authors would like to thank PEDECIBA (MEC, UdelaR, Uruguay).

%%%%%%%%%%%%%%%%%%%%%%%%%%%%%%%%%%%%%%%%

\section*{\label{sec:data} Data availability}

The data that support the findings of this study are available from the corresponding author upon reasonable request.

%%%%%%%%%%%%%%%%%%%%%%%%%%%%%%%%%%%%%%%%

\nocite{*}
\bibliography{biblio_fountainAction}
%\bibliography{aipsamp}% Produces the bibliography via BibTeX.

\end{document}